\documentstyle{article}
\newtheorem{th}{Theorem} 
\newtheorem{lm}{Lemma}
\newtheorem{df}{Definition}  
\newtheorem{pr}{Proposition}
\newcommand{\bth}{\begin{th}}
\newcommand{\eth}{\end{th}}
\newcommand{\blm}{\begin{lm}}    
\newcommand{\elm}{\end{lm}}
\newcommand{\bdf}{\begin{df}}   
\newcommand{\edf}{\end{df}} 
\newcommand{\bpr}{\begin{pr}}
\newcommand{\epr}{\end{pr}}
\newcommand{\bpf}{\noindent {\bf Proof:} }
\newcommand{\epf}{$\bullet$\par\vspace{1.8mm}\noindent}
\newcommand{\beq}{\begin{equation}}  
\newcommand{\eeq}{\end{equation}\par\noindent}
\newcommand{\beqa}{\begin{eqnarray*}} 
\newcommand{\eeqa}{\end{eqnarray*}\par\noindent}
\newcommand{\beqn}{\begin{eqnarray}} 
\newcommand{\eeqn}{\end{eqnarray}\par\noindent}  
 
\newfam\msbfam
\font\tenmsb=msbm10                     \textfont\msbfam=\tenmsb
\font\sevenmsb=msbm7            \scriptfont\msbfam=\sevenmsb
\font\fivemsb=msbm5                     \scriptscriptfont\msbfam=\fivemsb
\def\Bbb{\fam\msbfam \tenmsb}

\title{\bf A Representation for Compound\\ 
Quantum Systems as Individual Entities:\\ 
Hard Acts of Creation and Hidden Correlations.}
\date{}
\author{Bob Coecke\thanks{Free University of Brussels (VUB), Department of Mathematics, Pleinlaan 2, B-1050
Brussels; bocoecke@vub.ac.be\,.}} 
\begin{document}
\maketitle
\vspace{-6.cm}
\noindent
Appeared in Foundations of Physics {\bf 28}, 1109--1135 (1998) --- submitted in 1996.   
\vspace{5.5cm}
\noindent
\begin{abstract}
\noindent  
We introduce an explicit definition for 'hidden correlations' on
individual entities in a compound system: when one individual entity is measured, this induces a
well-defined transition of the 'proper state' of the other individual entities.  We prove that every
compound quantum system described in the tensor product of a finite number of Hilbert
spaces can be uniquely represented as a collection of individual(ized) (peudo-)entities between which
there exist such hidden correlations.  We investigate the significance of these hidden correlation
representations within the so-called ``creation-discovery-approach'' and in particular their
compatibility with the ``hidden measurement formalism''.  This leads us to the introduction of the
notions of 'soft' and 'hard' 'acts of creation' and to the observation that our approach can be seen
as a theory of (pseudo-)individuals when compared to the standard quantum theory. (For a presentation
of some of the ideas proposed in this paper within a quantum logical setting, yielding a structural
theorem for the representation of a compound quantum system in terms of the Hilbert space tensor
product, we refer to \cite{coe2000}.)    
\end{abstract} 

  
\section{Introduction.}         

Already for some time the study of compound systems has been one of the main issues
of the foundations of quantum physics, studied in great detail by many authors (see for example
\cite{bell64}, \cite{cla}, \cite{esp}, \cite{free72}, \cite{schro35} and
\cite{schro36}).  Nevertheless, to the present author's knowledge, all literature before the 80's
only refers to compound quantum systems in so called 'entangled states', i.e., compound systems
described in the tensor product of two Hilbert spaces.  In 1981 Aerts showed that separated quantum
entities cannot be described by the property lattice of a tensor product 
(see \cite{aer81}, \cite{bel} and
\cite{fra}).  As a
consequence, alternative ways of describing compound systems should be considered.\footnote{See also
\cite{coe2000} and \cite{val2000} for recent perspectives.} In particular, one should try to find a scheme for
the representation of compound systems in which as well the separated as the non-separated case fit in a
conceptually consistent way.  Such an attempt has been made in
\cite{aer91} and in \cite{coe96}: the entities in the compound system are considered
as {\it individual(ized) (pseudo-)entities} between which there exist a
specific kind of correlations, called {\it correlations of the second kind}.  These correlations of
the second kind were defined by Aerts in
\cite{aer91} as {\it correlations which were not present before the measurement but that are 
created during and by the measurement process}.  
In fact, as we show in the next section, this specific approach towards compound systems can
be considered as an aspect of a more general approach, namely the so-called creation-discovery approach,
discussed in \cite{aer97} and
\cite{aercoe} (a brief outline of this approach can be found in the next
section).  As Aerts shows in
\cite{aer81}, it is the presence of a non-empty {\it act of creation}, formalized as correlations of
the second kind, which is responsible for the violation of Bell-like inequalities (what we precisely
mean by this 'act of creation' gets clear in the next sections).   In the case that the individual
entities in the compound system are initially separated, we will consider this act of creation as {\it
empty}.  Thus, this concept of correlations of the second kind delivers a tool to consider
'entangled' and 'separated' entities within one approach.  However, the approaches introduced in
\cite{aer91} and in
\cite{coe96} are essentially metaphorical.
Moreover, the model systems that have been introduced only relate to a few compound
quantum systems with a description in 
${\Bbb C}^2\otimes{\Bbb C}^2$ or ${\Bbb C}^3$, and although all
these model systems have from a structural point of view a similar kind of
correlations of the second kind, an explicit mathematical definition has never
been given.   
  
In this paper we state a more explicit definition for correlations of the
second kind, which is fulfilled by the model systems of \cite{aer91} and
\cite{coe96}, and which enables a generalization beyond the Hilbert space tensor product
representation used in quantum theory, in the sense that we are not bound to the Hilbert space
tensor product for the description of compound systems, nor are we bound to the Hilbert space
itself for the representation of the states and the properties of the entities.  However, in order to
obtain the right results for the specific case of pure quantum theory, we prove that every compound
quantum system described in the tensor product of a finite number of Hilbert spaces has a {\it unique}
representation as a collection of individual entities on which we introduce our correlations of the
second kind.  As a consequence, the approach introduced in this paper generalizes the
quantum description for compound systems.   

The general conceptual nature of the representation 
introduced in this paper will be discussed in detail in the following section, but we
already mention some of it in a brief way. Following Piron's conception of {\it state} we consider the
state of a quantum system as a {\it complete} representation of the entity's elements of reality,
i.e., the state represents all actual properties of it, where the collection of all properties is
defined as the equivalence classes of the collection of all possible {\it questions} corresponding
with an explicit experimental procedure, and where {\it actuality} means that we obtain a positive
answer with certainty when the procedure is performed (see 
\cite{aer81}, \cite{moo} and \cite{pir}).\footnote{The present author is aware of the discussions,
i.e., the criticisms and replies, which are still running  on this approach to physics.  For an
overview on this debate we refer to \cite{bit82}, \cite{cat91}, \cite{cat93}, \cite{fou84} and
\cite{moo97}.  However, since in the past most criticism seems to be due to misunderstandings (see 
\cite{cat91}, \cite{fou84} and \cite{moo97}),  
we do choose to apply the conception of a state as it occurs within this approach.}  As a 
consequence, with 'state' we mean 'pure state': the  mixed states that occur in standard quantum
theory should be considered as due to a lack of knowledge on the state of the entity.  Since
this point of view works very well (read, {\it is compatible}) with the creation-discovery approach
(see \cite{aer94}), we take it as a starting point.  Since in this paper, we want to consider
individual entities within the compound system, we will be forced to extend this conception of
state.  This extension of Piron's concept of a state will be called the {\it proper state} of an
{\it individual entity} within a {\it compound system}.  
  
In our definition of correlations
of the second kind we require a {\it deterministic dependence} of the created correlations on the
transition of the proper state of the individual entity which is submitted to a measurement. 
As a consequence, this approach to compound
systems can be considered as conceptually compatible with another ingredient of the
creation-discovery approach, namely the {\it hidden measurement formalism} for quantum
measurements\footnote{This idea of hidden measurements has been introduced by Aerts in
\cite{aer86}, it has been developed into a mathematical framework in
\cite{aer94}, \cite{coe95} and \cite{coe96b}, and constitutes a fundamental aspect
of the creation-discovery approach (see \cite{aer97} and
\cite{aercoe}).} which will be discussed in the next section.  Therefore we call the specific kind of
correlations of the second kind that we define in this paper {\it hidden correlations}. 
In section \ref{comb.HM.HC} we formally combine hidden measurement representations and hidden
correlations. From this will follow that our approach can be considered as a 'theory
of individuals', compared to the standard treatment of compound systems in quantum theory.    
  
The mathematical core of the construction that we make for the case of quantum theory is essentially
based on Schmidt's  biorthogonal decomposition lemma (see \cite{schm}), which has been used by von
Neumann to explain how physical quantities of the subsystems of compound quantum systems are related
(see
\cite{neu}).\footnote{For the specific case of the tensor product of two Hilbert spaces, there is
some similarity with Herbut and Vuji\~ci\'c's representation for two spatially separated correlated
systems by two density matrices and a correlation operator (introduced and studied in 
\cite{her1} and \cite{her2}).}   We want to remark that the use of the biorthogonal
decomposition within our construction is not of the same fundamental nature as it is the case for the
modal interpretations of quantum mechanics (see for example \cite{hea89} and \cite{fra}).  In fact,
in our case it could even be avoided in the main formulation of the representation, but this causes
much more complicated expressions.  Moreover, the uniqueness of the representation is due to the
existence of the 'almost' uniqueness of the biorthogonal decomposition\footnote{I would like to
thank Dennis Dieks and Pieter Vermaas for pointing out the different nature of the biorthogonal
decomposition in our construction and in the modal interpretation in an explicit way.  The kernel of
this fundamental difference is the fact that "the distinguishing feature of the modal
interpretations of quantum mechanics is their abandonment of the orthodox eigenstate-eigenvalue
rule" (see \cite{cli95}) in contrary to our representation where Pauli's definition of a measurement
of the first kind (see
\cite{pau} and \cite{pir}), a generalization of von Neumann's eigenstate-eigenvalue
rule, plays a crucial role as a formalization of the explicit state transitions which we consider
as an essential ingredient of a measurement (a more detailed explanation of this matter is contained
in the following sections of this paper).}. 
 
Finally we want to remark that our approach/representation for obtaining all possible tensor
product states through 'hidden correlations' has also from a purely formal point of view some
interesting extra features relative to for example Herbut and Vuji\~ci\'c's representation, due to
the generalization to more than two Hilbert spaces, and due to the very natural way in which we
obtain uniqueness.   Thus, we think that the results presented in this paper contain some new formal
ingredients, even in absence of the extended creation-discovery approach. A first application
of these ingredients can be found in
\cite{spinN}, where we apply the formal results of this paper in order to construct a representation
for a spin-$S$ entity as a compound system consisting of $2S$ individual spin-$1/2$
entities between which there exist hidden correlations. 

\section{The creation-discovery approach and compound systems.}

In this section we introduce the general philosophical framework, called the creation-discovery
approach, that will be considered as the general discourse and mode of thinking in this paper.  As
already remarked in the previous section, some formal aspects of the representation could be
considered without the framework presented in this section.  However, the embedment of the
representation within a clear and global approach accents the motivation behind 
it. We also are aware of the fact
that the philosophical framework presented in this section requires much more pages of
explanation for really deserving the name of a 'philosophical framework'.  Nonetheless, we think that
this more or less intuitive presentation preserves a sufficient explanation of our
motivation.    

\subsection{The creation-discovery approach.}

D. Aerts proposed in the 80's a
way to identify the
physical aspects that are at the origin of the structural differences between quantum and classical
theories (see \cite{aer81} and \cite{aer86}). It are mainly two aspects that determine these
structural differences, in the sense that we obtain
quantum-like probability structures if the measurements needed to test the properties of the system
are such that:
\par
\smallskip 
\noindent
{\it{\bf 1.} The measurements are not just observations but provoke a real change of the
state of the system. {\bf 2.} There exists a lack of knowledge on what precisely happens during the 
measurement process.}
\par
\smallskip
\noindent
The first aspect, the change of state, can be interpreted as an 'act of creation' on the entity
under study. It is indeed the external device that provokes the change of state during the
interaction with the entity.  If there is not such a change of state, we call the measurement a
discovery.  The second aspect, the presence of the lack of knowledge on the
precise act of creation which results from an interaction with the measurement context, lies at the
origin of the so called indeterministic nature of quantum measurements and can be formalized as a
lack of knowledge on the precise measurement that is actually performed:   
\par 
\smallskip
\noindent
{\it{\bf 1.} With each real measurement $e$
corresponds a collection of deterministic measurements $e_\lambda$, called 'hidden measurements'.
{\bf 2.} When a measurement $e$ is performed on an entity in a state $p$, then one of the  hidden
measurements $e_{\lambda}$ takes place. The probability finds its origin in the lack of knowledge
about which one of the hidden measurements effectively takes place\footnote{As is shown in \cite{aer86}
and \cite{bofr}, these hidden measurement representations are not in contradiction with the
NoGo-theorems about hidden variables (all inspired by the von Neumann proof in \cite{neu}) since the
variables in the hidden measurement approach are contextual by definition.  For more details on this
hidden measurement approach we refer to \cite{aer86},
\cite{aer94},
\cite{aer97}, \cite{aercoe}, \cite{coe95} and \cite{coe96b}.  Another model system in which we
encounter an introduction of parameters representative for this kind of lack of knowledge
situation is the Gisin-Piron model in
\cite{gis81}.  However, this aspect of the model was not the main topic of their paper.  Gisin and
Piron mainly wanted to show that it is possible to find a dynamical equation for a 'state transition'
during a measurement, i.e., a collapse without going through a mixture.}.} 
\par
\smallskip
\noindent
In \cite{aer86} and \cite{aer94} it is proved that such a representation exists for quantum measurements
described in finite dimensional Hilbert spaces, and a proof for the infinite dimensional case is
delivered in \cite{coe95} and \cite{coe96b}. It is important to remark that in this
hidden measurement formalism:   
\par
\smallskip
\noindent
{\it The state $p$ does not depend on the parameter
$\lambda$ and the selection of $\lambda$ is also independent of the state $p$\footnote{In fact,
in a somewhat modified version of the hidden measurement formalism elaborated in
\cite{bofr}, this is not true anymore.  As a consequence, we loose the
clean cut between 'measurement-dependence' and 'entity-dependence' of relevant parameters in a formal
description of a measurement.  This altered version of the original hidden measurement approach has
the great advantage that it takes in an a priori way the property structure of the entity into
account, which leads to some remarkable uniqueness theorems.  Unfortunately, one
looses the extreme philosophical simplicity of the original hidden measurement approach.  We also
want to remark that for all results presented in this paper, and also for the presented conceptual
framework in relation to the description of compound systems, it wouldn't pose any additional problem
if we would replace the hidden measurement ingredient by this recently introduced alternative for
it.  However, we would loose some of the transparency obtained in the resulting models, and therefore
we choose for the purpose of this paper to stick to the original hidden measurement formalism.}.}  
\par
\smallskip
\noindent
The formalism has been generalized beyond the quantum framework in \cite{aer94}, \cite{coe95} and
\cite{coe96b}. 
We briefly present the hidden measurement formalism following the definitions of
\cite{coe96b}. Let
$\Sigma$ be a collection of states,
$\Lambda$ the collection of parameters for the hidden measurements in a hidden measurement representation and
let $\mu:{\cal B}(\Lambda)\rightarrow [0,1]$, with ${\cal B}(\Lambda)$ a
$\sigma$-field of subsets of
$\Lambda$, determine the relative frequency of occurrence of $\lambda\in\Lambda$.
For the purpose of this paper, we suppose that due to the state transition aspect of a measurement
$e$, the outcomes of this measurement can be identified 
in a one to one way by a collection of outcome
states
$\Sigma_e$.
For a fixed $\lambda\in\Lambda$, the measurement process is strictly classical: 
for every {\it hidden measurement} $e_\lambda$ there exists a {\it strictly classical
observable}  
$\varphi_\lambda:\Sigma\rightarrow \Sigma_e$. 
Thus we can represent the {\it unknown but relevant content of the measurement interaction} for the
measurement process as a couple consisting of a set of strictly classical
observables  
and a probability measure:
\beq
\left\{ 
\begin{array}{l}
\Phi=\{\varphi_\lambda:\Sigma\rightarrow \Sigma_e|\lambda\in\Lambda\}\\
\mu:{\cal B}(\Lambda)\rightarrow [0,1] 
\end{array} 
\right\delimiter0
\eeq
For every
possible initial state $p\in\Sigma$, a measurement $e$ is characterized by a 
probability measure $P_{p,e}:{\cal B}(\Sigma_e)\rightarrow [0,1]$ (${\cal
B}(\Sigma_e)$ is a $\sigma$-field of subsets of $\Sigma_e$).  Also every hidden measurement
$e_\lambda$ corresponds with a
probability measure, namely $P_{p,\lambda}:{\cal
B}(\Sigma_e)\rightarrow\{0,1\}$ which is such that
\beq
P_{p,\lambda}(\{q\})=1\Leftrightarrow\varphi_\lambda(p)=q
\eeq
Thus we can relate the above introduced probability measures\footnote{To assure the existence of
the following expressions all arguments of probability measures need to be measurable.  A necessary
and sufficient condition to achieve this can be found in
\cite{coe96b}.}: 
\beq
P_{p,e}(q)=\mu\bigl(\{\lambda|P_{p,\lambda}(\{q\})=1\}\bigr)
=\mu\bigl(\{\lambda|\varphi_\lambda(p)=q\}\bigr)
\eeq
All this has led to a general axiomatics for context dependence and a classification of
all possible hidden measurement representations (see \cite{coe96b}). From this classification 
it follows that in general, the hidden measurement approach delivers no {\it a priori} way to
represent the 'true structure' of a physical entity (see
\cite{coe96b} and \cite{bofr}).  In fact, it is one of the main aims of this paper to introduce a
way to describe {\it compoundness} of physical systems within this very general approach.  More
precisely, the hidden measurement representations for the composing entities together with the hidden
correlations deliver a new hidden measurement representation for the compound system which takes
into account the true nature of the compound quantum entity, i.e., the compoundness itself.  As
already mentioned  in the introduction, this specific 'compatibility', or even better
'complementarity', is formally elaborated in section 3 of this paper. 

Although we are always able to formalize quantum-like probability structures as a lack of knowledge
on the measurement that {\it will} take place, this does not exclude the occurrence of lack of
knowledge on the initial state (for example: an entity in a statistical ensemble in which appear
different states).  In such a case, it is more natural to formalize this 'lack of knowledge on the
initial state'-aspect by a probability measure defined on measurable subsets of the state space. 
For the case of quantum theory, the combination of the probabilistic behavior in a
measurement for a well-defined initial state (i.e., a
positive normalized measure defined on the subspaces of the Hilbert space which is additive on every
countable subset of mutual orthogonal subspaces) and   
a lack of knowledge on the initial state (i.e., a probability measure defined on the set of
states) corresponds due to Gleason's theorem in a one to one way with a
density matrix, i.e., with a positive operator on the Hilbert space of unit trace (see
\cite{glea}).  

However, although formally we have a situation of a probability measure on the set of states and a
probability measure on the set of hidden measurements, there is a very important conceptual difference
between the two kinds of uncertainty that correspond with them.  Namely, at the initial stage of an
experimental setup, the entity {\it is} in a state, we only don't know which one.  On the
contrary, one cannot speak about an initial unknown hidden measurement, but only about a
formalization of the interaction that {\it will} take place when we actually decide to
perform the measurement, i.e., if there is no measurement, the parameters $\lambda$ that
characterize the possible formalizations of the deterministic measurement process through hidden
measurements are meaningless.  As a consequence, the word 'hidden' represents in the
creation-discovery approach a combination of an aspect of potentiality (i.e., what might become
but what is not) combined with an aspect of uncertainty\footnote{This new status that we attach to
the word 'hidden' definitely differs from its historical origin, namely the hidden variable
disputes.  Therefore one might think that for our purpose, the word hidden might be badly
chosen.  However, according to Aerts, at the time he introduced his first hidden measurement
model, the living paradigms and the not yet sufficiently  refined ideas he had on the hidden
measurements did justify that choice of the name.}.  As we will see in the following
sections, this same point of view applies to the hidden correlations.      

\subsection{Soft and hard acts of creation.}

For the purpose of this paper, we need to refine the notion of act of creation.   In the previous
subsection we have identified the change of state that occurs during a measurement as an 'act of
creation'. In the case of a pure quantum measurement, this change of state corresponds with a unitary
(read 'structure preserving') transition.  Thus, the act of measurement doesn't imply a change of the
set of properties, and thus, also not a change of the set of states, but only a change of the actual
properties.  However, one could imagine some more drastic acts of
creation.  Consider for example a uniform sphere.  The states of such a sphere correspond with
possible spatial coordinates of its center, and a state transition corresponds with a translation.
Suppose now that one cuts up the sphere in two halve spheres.  Every one of these halve spheres has
now, apart from their translational properties, also angular properties.  Thus, due to this cutting,
new properties occur, and thus, we obtain a set of states that is different from the original one. 
Moreover, even if we keep the two halve spheres together, and consider them as one entity, the
presence of the cut yields a different set of states than the one for the uniform sphere.  Thus, by
'cutting' we perform an 'act of creation': we create new properties.  In fact, since a lot of authors
define an entity by a set of states, one might say that we create new entities.  To distinguish between these two different acts of creation, respectively the one
provoking a change of state and the one provoking a change of the set of states, we will call them
respectively {\it soft} and {\it hard} acts of creation.  Another example is an Aspect-like
measurement, where one destroys the entanglement and creates new properties, which refer to the
polarization angles, in order to obtain two separate photons.   Nonetheless, as shown in
\cite{aer91}, in this case we also need a soft act of creation in order to obtain the required quantum
structure.    Thus, in general we will have to consider both kinds of acts of creation together.

We will now analyze the change of the state space due to the hard act of creation for the case of
two individual spin-1/2 entities described by the so called singlet state, an example which is
conceptually almost identical to an aspect-like experiment and which was also considered by
Aerts in
\cite{aer91}.  Before we perform the measurement, the angular properties are traditionally described
by the singlet state.  In fact, from the experimental data that result from measurements on it, it
follows that 'being in a singlet state' means nothing else than 'having no angular properties at
all', since for every possible measurement referring to spin we have a uniform probability
distribution for the possible outcomes (see
\cite{aer91}).  After a measurement we obtain two separated spin-1/2 entities, in standard quantum
mechanics described as products in the tensor product, but which might of course as well be described
by a cartesian product of two spin-1/2 state spaces.  Thus we have a change of the states space from
a singleton to a cartesian product, due to the hard act of creation.  The power of the standard
quantum description lies in the fact that within one representation space, the 'large' space
${\Bbb C}^2\otimes{\Bbb C}^2$, we are able to represent both state spaces relatively to
each other.  However, this requires the introduction of superselection rules to forbid the use of
'meaningless' elements in ${\Bbb C}^2\otimes{\Bbb C}^2$. Unfortunately, when we start to
consider measurements on compound quantum systems consisting of more than two individual entities
things get more complicated.  This is discussed in the next section.
  
\subsection{Individual entities and proper states.}

As already mentioned in the introduction, if one wants to
talk about the elements of reality of an individual entity within a compound system, the concept of a
{\it state} of \cite{aer81}, \cite{aer94}, \cite{moo} and \cite{pir} should be extended (read {\it
weakened} or {\it fuzzyfied}) beyond its explicit operational definition.  
We first consider the example of an Aspect-like
experiment or equivalently, the case of
two individual spin-1/2 entities described by the so called singlet state: before a measurement on
one of the individual entities has been performed, there exist no properties in Piron's sense
referring to {\it spin}, and thus, strictly spoken, nothing can be said on it; in this case of two
individual spin-${1\over 2}$ entities described by the so called {\it singlet state} this causes not
too much problems since before the actual act of creation took place there are no spin 
{\it tendencies}\footnote{Since in this subsection we are discussing 
Piron's concept of 'state', which is itself based on a very rigorous notion of 'property', we will
from now on avoid the use of the word property in its more intuitive sense where it refers to a
certain tendency of a system in a measurement, by using the word 'tendency' itself.} at all (in the
same sense as the absence of angular tendencies for a uniform sphere) and after a
measurement on one of the individual entities these entities are separated such that both have taken a
spin-${1\over 2}$ state; however, as it will follow from the results that we obtain in section 4 of
this paper, when we consider three or more individual spin-${1\over 2}$ entities in a compound
system, after a measurement on one of the individual entities, the two not yet measured
entities do have tendencies towards certain states, although they do not have a probabilistic
behavior that can be explained by a state, i.e., we are in a kind of {\it intermediate situation}
between {\it no spin-${1\over 2}$ tendencies} and {\it a spin-${1\over 2}$ state}. A second example
is the one of a spin-$S$ entity when it is considered as a compound system consisting of $2S$
individual spin-${1\over 2}$ entities (this point of view is introduced and studied in
\cite{spinN}): in this case it seems to be impossible to measure the individual entities one by one in
a controlled way, and thus, there seem to exist no experimental procedures that test the properties of
the individual entities, which implies that we cannot speak about properties and thus, also not
about states.   Therefore we decided to speak about the {\it proper state} of an
individual entity.  Consequently, following the above mentioned point of view that an entity is 
defined by a set of states, the concept of an {\it individual entity} itself is again an extension of
the concept of an entity.  Obviously, a lot more could be said or investigated on these two notions
proper state and individual entity.  However, this would take us too far away from our main goal, so
for the purpose of this paper we limit ourselves to a minimal intuitive introduction of these
concepts in a sufficient way to avoid confusion or misunderstandings, leaving explicitation or
refinement for future writings, discussions and investigations.   Nonetheless, we do have to
introduce a formal representation for the 'relevant content' of these proper states.   Since in this
paper we will consider measurements on individual entities which are such that after the measurement
the individual entity is separated from the others in the compound system, i.e., the individual
entity becomes an entity, we can consider the set of states of it.  As a consequence we could decide
to represent the proper state relative to these states, by considering the probabilities for a proper
state transition to these states in a measurement.  Of course, there is no a priori reason why such a
representation would be identical if we consider different measurements with some common possible
state transitions, and therefore, we are bound to include the different possible measurements in our
representation, i.e., for every possible measurement on the individual entity in a proper state we
should give a probability measure defined on the possible states for the entity after the
measurement.  In the case of individual quantum entities we will see that that this complex
representation reduces into one formal object, namely a density matrix (see section
\ref{sect.HC.QM}).  However, it is clear that within our creation-discovery approach there is a
definite conceptual difference between this use of the density matrix as a representation for proper
states and its usual use (i.e., a representation for a measurement that formalizes a lack of
knowledge on the initial state combined with one on the state transition that will take place when we
perform the measurement): in the case of a proper state it makes no sense anymore to consider a lack
of knowledge on the state, since these states only become relevant characterizations of the individual
entity after the hard act of creation that happens during the measurement\footnote{In fact, in
accordance with the meaning that we have given to the word hidden in 'hidden measurement', we might
even call these states 'hidden states': they formalize what might become when we perform a
measurement, i.e., when we perform an act of creation.  However, since this expression 'hidden state'
is explicitly used in the hidden variable context it is better to avoid the use of it in this
paper.}.         
 
\subsection{Hidden correlations.}

Let $S$ be a compound  system consisting of a finite collection of individual entities $\{S_\nu\}_\nu$
(with this  notation we refer to a set indexed over all $\nu$; the  symbol $\nu$ will consequently be
used as a parameter that takes
$\alpha,\beta,\gamma,\ldots,\kappa,\lambda,\mu,\ldots,\xi,\upsilon,\zeta$, i.e., the labels of
individual entities in the compound system, as values).  
Every individual entity $S_\alpha\in\{S_\nu\}_\nu$ has a set of proper states represented by the
set $\bar\Sigma_\alpha$ and a set
$\Sigma_\alpha$ referring to the states the individual entity can take when it is separated from
the other ones.   Due to the state transition aspect of a soft act of creation, we follow Pauli's
definition of a measurement of the first kind (see \cite{pau} and \cite{pir}), i.e., we suppose that
after a measurement on an entity, its state has changed into a state that corresponds with
the obtained outcome, and we call this state an {\it outcome state}. 
Thus, a measurement on $S_\alpha$ which is such that after the measurement the individual entities
in the compound system are separated, is characterized by a set of outcome states
$\{\phi_{\alpha,i}\}_i$ (with this notation we refer to a set indexed over all 'relevant' indices
$i$; what we mean by the relevant indices will follow from the context where we apply this notation).
In fact, in this paper we will always  consider such measurements\footnote{In \cite{spinN} we
consider measurements that do not separate the individual entities. However,  the 'proper outcome
states' can be 'represented' in a one to one way by 'states', and thus the outcomes of a measurement
on the compound systems can be written as a product
$\times_\nu\phi_\nu$.  As a consequence, the obtained results in this paper can be applied in 
\cite{spinN}.}.  If we consider an obtained outcome
state in a measurement on $S_\alpha$ we denote it as $\phi_\alpha$. 
A measurement on the compound system $S$ corresponds with one measurement on every $S_\alpha$, and an
outcome of this measurement can be written as a cartesian product
$\times_\nu\phi_\nu$.    
\bdf
We say that there exist 'hidden correlations' between individual entities in a collection
$\{S_\nu\}_\nu$ if: 
{\bf 1.} a measurement on one individual entity $S_\alpha$ induces a change of the proper state of the
other individual entities
{\bf 2.} this change of proper state depends deterministically on the proper state transition of
$S_\alpha$ 
{\bf 3.} after this measurement, the state of
$S_\alpha$ cannot be influenced anymore by measurements on other entities.
\edf 
It is clear that the use of the word 'hidden' is justified by the fact that the specific correlation
that occurs depends on the effect that the act of creation of the measurement has on the proper
state of the individual entity $S_\alpha$.  In the next section we present a representation that
fulfills the above stated definition for the case of the quantum formalism.   For every measurement
on an individual entity $S_\alpha$, we have to define one map for every other individual
entity $S_\beta$ that characterizes the induced change of proper state.  In the specific
representation that we introduce in the following section, these maps only depend in an explicit way
on the outcome state of the first measurement.  Thus, we will have to define: 
\beq
f_{\beta\circ\alpha}:\Sigma_\alpha\rightarrow\bar\Sigma_\beta:
\phi_\alpha\mapsto\omega_{\beta\circ\alpha} 
\eeq
which determines a state transition of $S_\beta$ from a proper state $\omega_\beta$ into a proper
state $\omega_{\beta\circ\alpha}$, due to a transition of the proper state of $S_\alpha$ from
$\omega_\alpha$ into an outcome state $\phi_\alpha$.  Analogously we define: 
\beq
f_{\mu\circ\lambda\circ\kappa\circ\ldots\circ\alpha}:\Sigma_\lambda\rightarrow\bar\Sigma_\mu:
\phi_\lambda\mapsto\omega_{\mu\circ\lambda\circ\kappa\circ\ldots\circ\alpha} 
\eeq
such that if due to a measurement on $S_\lambda$ after we have already performed
measurements on 
$S_\alpha,\ldots,S_\kappa$, the proper state of
$S_\lambda$ becomes
$\phi_\lambda$, the proper state of every not yet measured individual entity $S_\mu$ changes to 
$\omega_{\mu\circ\lambda\circ\kappa\circ\ldots\circ\alpha}=
f_{\mu\circ\lambda\circ\kappa\circ\ldots\circ\alpha}(\phi_\lambda)$.
Of course, in general one could consider more general kinds of
$f_{\mu\circ\lambda\circ\kappa\circ\ldots\circ\alpha}$ in the sense that they do not only depend on
$\phi_\lambda$ but also on $\omega_{\mu\circ\kappa\circ\ldots\circ\alpha}$,
on $\omega_{\lambda\circ\kappa\circ\ldots\circ\alpha}$, or even on an additional variable. However, in
theorem
\ref{th.correct} in section
\ref{sect.HC.QM}  we will prove that this specific kind of dependence on the state transition covers
all possible hidden correlation representations for compound quantum systems described in a tensor
product.  Also for the purpose of showing some peculiar features of our approach in contrary to the
standard quantum approach, these kinds of hidden correlations suffice.  We remark that one can consider
the change of state of the individual entities (caused by a measurement on one of them) as due
to the presence of a 'preserved quantity' (such as momentum in classical physics), in the sense
that the change of state of the measured individual entity ('forced' by the act of
creation in the measurement on it) is compensated by the change of state of the other
individual entities. The deterministic dependence represented by the map
$f_{\mu\circ\lambda\circ\kappa\circ\ldots\circ\alpha}$ is an obvious assumption since the 'preserved
quantity' is an intrinsic quantity of the collection of entities, and thus, these changes of state are
not due to an act of creation during interaction with an external context.  

\section{Combining hidden measurements and hidden correlations.}\label{comb.HM.HC}

Suppose that we have a given hidden measurement representation for the measurements on the
individual entities in a compound system $S$, i.e., for all $\alpha$ there exists: 
\beq
\left\{ 
\begin{array}{l}
\Phi_\alpha=\{\varphi_{\alpha,\lambda_\alpha}:\bar\Sigma_\alpha\rightarrow\{\phi_{\alpha,i}\}_i|
\lambda_\alpha\in\Lambda_\alpha\}\\
\mu_\alpha:{\cal B}(\Lambda_\alpha)\rightarrow [0,1]
\end{array} 
\right\delimiter0
\eeq
and suppose that we also have a hidden correlation representation which formalizes the
specific compoundness of the system.  Then we can define a new hidden measurement representation for
the compound system $S$.  We represent the proper states of the individual systems in a cartesian
product, i.e., $\Sigma_S=\times_\alpha\bar\Sigma_\alpha$, and we also define
$\Lambda_S=\times_\alpha\Lambda_\alpha$. For a measurement on the individual entities according
to the ordering $\alpha,\beta,\ldots,\upsilon,\zeta$ we define: 
\beq\label{Eq.varph.HM+HC}
\varphi_{S,\lambda}:\Sigma_S\rightarrow\times_\alpha\{\phi_{\alpha,i}\}_i:
\times_\alpha\omega_\alpha
\mapsto
\left( 
\begin{array}{l}
\hspace{3.3cm}\varphi_{\alpha,\lambda_\alpha}(\omega_\alpha)\\ 
\hspace{2.3cm}\varphi_{\beta,\lambda_\beta}\circ
f_{\beta\circ\alpha}\circ\varphi_{\alpha,\lambda_\alpha}(\omega_\alpha)\\ 
\hspace{3.8cm}\ldots\\
\varphi_{\zeta,\lambda_\zeta}\circ
f_{\zeta\circ\upsilon\circ\ldots\circ\beta\circ\alpha}\circ\varphi_{\upsilon,\lambda_\upsilon} 
\circ\ldots\circ
\varphi_{\beta,\lambda_\beta}\circ
f_{\beta\circ\alpha}
\circ\varphi_{\alpha,\lambda_\alpha}(\omega_\alpha) 
\end{array} 
\right)
\eeq
Following the above stated definitions we have:
\beq
\phi_{\alpha}=\varphi_{\alpha,\lambda_\alpha}(\omega_\alpha)
\eeq
Since we have
$\omega_\beta=f_{\beta\circ\alpha}(\phi_{\alpha})
=f_{\beta\circ\alpha}\circ\varphi_{\alpha,\lambda_\alpha}(\omega_\alpha)$ and since
$\phi_\beta=\varphi_{\beta,\lambda_\beta}(\omega_\beta)$ we also have: 
\beq
\phi_\beta=\varphi_{\beta,\lambda_\beta}\circ
f_{\beta\circ\alpha}\circ\varphi_{\alpha,\lambda_\alpha}(\omega_\alpha)
\eeq
By an analogous induction on the number of consecutively performed measurements on the individual
entities we obtain:
\beq
\phi_{\zeta}=\varphi_{\zeta,\lambda_\zeta}\circ
f_{\zeta\circ\upsilon\circ\ldots\circ\beta\circ\alpha}\circ\varphi_{\upsilon,\lambda_\upsilon} 
\circ\ldots\circ\varphi_{\beta,\lambda_\beta}\circ f_{\beta\circ\alpha}
\circ\varphi_{\alpha,\lambda_\alpha}(\omega_\alpha)
\eeq
As a consequence, eq.(\ref{Eq.varph.HM+HC}) displays the outcome of a measurement on $S$ up to a
value in $\Lambda_S$, by taking into account the transitions of proper states of the individual
entities that happen 'between' the measurements on the individual entities.  Thus, we obtain a new
hidden measurement representation: 
\beq\label{Eq.HM+HC}
\left\{ 
\begin{array}{l}
\Phi_S=\{\varphi_{S,\lambda}|\lambda\in\Lambda_S\}\\
\mu_S:{\cal B}(\Lambda_S)\rightarrow [0,1]
\end{array} 
\right\delimiter0
\eeq
which is such that $\times_\alpha\mu_\alpha$ is equal to $\mu_S$ for these
subsets of $\Lambda_S$ where both are defined. 
\par
\smallskip 
\noindent 
{\it We remark that in this new hidden measurement representation, the formal ingredient that
represents the specific kind of compoundness of the system is included in $\Phi_S$ and not in
$\Sigma_S$.  
Thus, in our representation for measurements on compound systems, we deal in same way with the
interaction of an individual entity with the other individual entities as with the interaction with
the measurement context.}   
\par
\smallskip
\noindent 
In the case of quantum mechanics, the formal ingredient that represents the kind of
compoundness is included in the so called 'state' of the compound system, i.e., in the representative
vector of the tensor product.  This observation allows us to say that {\it our approach can be
considered as a theory of individuals, compared to the standard treatment of compound systems in
quantum theory}.  In general, the representation of eq.(\ref{Eq.varph.HM+HC}) and eq.(\ref{Eq.HM+HC})
depends on the order of the measurements.  Since this order in which the measurements on the
individual entities are performed is definitely an ingredient of the 'context' of the individual
entities and not of the individual entities or even the compound system itself, this fact seems to
be very obvious.  Thus, when we consider standard quantum theory, the description of the compound
system seems to include an ingredient that
represents the order of the measurements on the individual entities, and this can at least be called
strange.  However, in this quantum case we have the exceptional situation that the obtained
probability structure does not depend on the order of the measurements on the individual entities,
such that we do not encounter equivalently strange phenomena on the formal level.  To conclude:
although this reasoning might be called somewhat intuitive, it definitely indicates that {\it in the
standard quantum theory there are serious problems when we want to consider individual entities
within a compound system}.      

\section{A hidden correlation representation for compound quantum systems described in the tensor
product of Hilbert spaces.}\label{sect.HC.QM}  

According to quantum mechanics, the states in $\Sigma_\alpha$ are described in a Hilbert space ${\cal
H}_\alpha$ and the compound system is described by $\Psi_S\in \otimes_\nu{\cal H}_\nu$. 
In this section we will show that it {\it requires} and {\it suffices} to consider the density
matrices as representations for the proper states\footnote{We remark that from a
geometric point of view, the density matrices can be seen as the convex closure of the
states.}, consequently denoted by
$\bar{\cal H}_\alpha$.  For reasons of simplicity we will refer to these density matrices as proper
states.   Also for reasons of simplicity we suppose that all measurements have a non-degenerate
discrete spectrum (this does not cause a loss of generality).     The outcome states for a
measurement on individual entities correspond with the eigenvectors of the self-adjoint operator that
represents the measurement, and an outcome of a measurement on the compound system $S$, which we have
denoted by
$\times_\nu\phi_\nu$, can now equivalently be represented as a vector in $\otimes_\nu{\cal H}_\nu$,
namely $\otimes_\nu\phi_\nu$.

\subsection{Explicitation of the representation.}

Let ${\cal H}$ and $\tilde{\cal H}$ be two Hilbert spaces.  
For every $\Psi\in {\cal H}\otimes\tilde{\cal H}$ there always exists a biorthogonal 
decomposition\footnote{In most of the cases, this biorthogonal decomposition is unique.  It is
possible to show that the representation that we are going to construct will not depend on the choice
of the decomposition for the cases that it is not unique. The proof of this can be found
in the appendix at the end of this paper.} such that we can write: 
\beq\label{eq:biorth.dec.}
\Psi=\sum_i\langle\psi_i\otimes\tilde{\psi}_i|\Psi\rangle\psi_i\otimes\tilde{\psi}_i 
\eeq
where $\{\psi_i\}_i$ and $\{\tilde{\psi}_i\}_i$ 
are two orthonormal sets of vectors respectively in ${\cal H}$ and $\tilde{\cal H}$.  For a proof we
refer to \cite{fra} or \cite{neu}.  
Clearly, $\{\psi_i\}_i$ (resp. $\{\tilde{\psi}_i\}_i$) can always be extended to a base of ${\cal H}$
(resp.
$\tilde{\cal H}$).  If no confusion is possible, we will apply the same notation  
$\{\psi_i\}_i$ (resp. $\{\tilde{\psi}_i\}_i$) for such a base of ${\cal H}$ (resp.
$\tilde{\cal H}$).  One easily verifies that for all 'extra' vectors that we introduce in order to
obtain a base we have that $\langle\psi_i\otimes\tilde{\psi}_i|\Psi\rangle$ is equal to zero or it
does not exist (this might happen in the case that ${\cal H}$ or $\tilde{\cal H}$ are not equal 
dimensional).  We'll assume that in eq.(\ref{eq:biorth.dec.}) the sum runs over all $i$ which are
such that $\langle\psi_i\otimes\tilde{\psi}_i|\Psi\rangle$ exists.
Let
$\Psi_S\in\otimes_\nu{\cal H}_\nu$.  For every $S_\alpha$, there exists
such a decomposition for $\Psi_S$ if we consider $\otimes_\nu{\cal H}_\nu$ as a tensor product of two
Hilbert spaces ${\cal H}_\alpha$ and $\otimes_{\nu\not=\alpha}{\cal H}_\nu$, i.e., we consider 
$\Psi_S\in{\cal H}_\alpha\otimes (\otimes_{\nu\not=\alpha}{\cal H}_\nu)$. In this case, we denote the
two orthonormal bases in the biorthogonal decomposition of $\Psi_S$ as
$\{\psi_{\alpha,i}\}_i$ (an orthonormal base for ${\cal H}_\alpha$) and
$\{\tilde{\psi}_{\alpha,i}\}_i$ (an orthonormal base for $\otimes_{\nu\not=\alpha}{\cal H}_\nu$). 
Given this decomposition, we define for every $S_\alpha$ a map:
\beq\label{eq.first.HC}
{\cal R}_\alpha:\otimes_\nu{\cal H}_\nu\rightarrow\bar{\cal H}_\alpha:\Psi\mapsto\omega_\alpha
\eeq
where $\omega_\alpha$ is the diagonal density matrix in the base
$\{\psi_{\alpha,i}\}_i$ which is such that the $i$th diagonal element is given by
$|\langle\psi_{\alpha,i}\otimes\tilde{\psi}_{\alpha,i}|\Psi\rangle|^2$. We define a map:
\beq
T_\alpha:{\cal H}_\alpha\rightarrow\otimes_{\nu\not=\alpha}{\cal H}_\nu:
\phi\mapsto
{1\over N(\phi)}
\sum_i\langle\psi_{\alpha,i}\otimes\tilde{\psi}_{\alpha,i}|\Psi_S\rangle
\langle\phi|\psi_{\alpha,i}\rangle\tilde{\psi}_{\alpha,i}
\eeq
where:
\beq
N(\phi)=\sqrt{\sum_i|\langle\psi_{\alpha,i}\otimes\tilde{\psi}_{\alpha,i}|\Psi_S\rangle|^2
|\langle\phi|\psi_{\alpha,i}\rangle|^2}
\eeq
We also define a map for every two individual entities $S_\alpha$ and $S_\beta$: 
\beq
{\cal R}_{\beta\circ\alpha}:\otimes_{\nu\not=\alpha}{\cal H}_\nu\rightarrow
\bar{\cal H}_\beta:\Psi\mapsto\omega_{\beta\circ\alpha}
\eeq
in an analogous way as ${\cal R}_{\alpha}$, but now by considering a biorthogonal
decomposition of $\Psi$ ($\in\otimes_{\nu\not=\alpha}{\cal H}_\nu$) in the tensor product ${\cal
H}_\beta\otimes (\otimes_{\nu\not=\alpha,\beta}{\cal H}_\nu)$ in stead of in ${\cal H}_\alpha\otimes
(\otimes_{\nu\not=\alpha}{\cal H}_\nu)$, i.e., 
if $\Psi=\sum_i\langle\psi_{\beta\circ\alpha,i}\otimes\tilde{\psi}_{\beta\circ\alpha,i}|
\Psi\rangle\psi_{\beta\circ\alpha,i}\otimes\tilde{\psi}_{\beta\circ\alpha,i}$ is again a biorthogonal
decomposition with $\{\psi_{\beta\circ\alpha,i}\}_i$ as a base for ${\cal H}_\beta$ and 
$\{\tilde{\psi}_{\beta\circ\alpha,i}\}_i$ as a base for 
$\otimes_{\nu\not=\alpha,\beta}{\cal H}_\nu$, then ${\cal
R}_{\beta\circ\alpha}(\Psi)=\omega_{\beta\circ\alpha}$ is the diagonal density matrix with 
$|\langle\psi_{\beta\circ\alpha,i}\otimes\tilde{\psi}_{\beta\circ\alpha,i}|\Psi\rangle|^2$ as
respective elements in the base $\{\tilde{\psi}_{\beta\circ\alpha,i}\}_i$.
Finally, we define:
\beq
f_{\beta\circ\alpha}:{\cal H}_\alpha\rightarrow
\bar{\cal H}_\beta
\eeq
such that:
\beq
f_{\beta\circ\alpha}={\cal R}_{\beta\circ\alpha}\circ T_\alpha
\eeq
If we perform a measurement on $S_\alpha$ and we obtain an outcome state $\phi_\alpha$,
then the proper state of $S_\beta$ (for all $\beta\not=\alpha$) changes to
$f_{\beta\circ\alpha}(\phi_\alpha)$.  
We still have to explain what happens in consecutive measurements. Suppose that after the
measurement on $S_\alpha$, we perform a measurement on $S_\beta$. We define a map: 
\beq\label{eq:Tba}
T_{\beta\circ\alpha}:{\cal H}_\beta\rightarrow\otimes_{\nu\not=\alpha,\beta}{\cal H}_\nu
\eeq
in analogy with $T_\alpha$, but now by using a
biorthogonal decomposition of 
the vector $T_\alpha(\phi_\alpha)$ in
${\cal H}_\beta\otimes(\otimes_{\nu\not=\alpha,\beta}{\cal H}_\nu)$ 
in stead of
the vector $\Psi_S$ in ${\cal H}_\alpha\otimes(\otimes_{\nu\not=\alpha}{\cal H}_\nu)$.  For every
third individual entity
$S_\gamma$ we define a map:
\beq
{\cal R}_{\gamma\circ\beta\circ\alpha}:\otimes_{\nu\not=\alpha,\beta}{\cal H}_\nu\rightarrow
\bar{\cal H}_\gamma:\Psi\mapsto\omega_{\gamma\circ\beta\circ\alpha}
\eeq
in an analogous way as ${\cal R}_{\beta\circ\alpha}$ and  ${\cal R}_\alpha$, but now by
considering biorthogonal decompositions in ${\cal H}_\gamma\otimes
(\otimes_{\nu\not=\alpha,\beta,\gamma}{\cal H}_\nu)$. Thus, we can define 
\beq
f_{\gamma\circ\beta\circ\alpha}:{\cal H}_\beta\rightarrow\bar{\cal H}_\gamma
\eeq
such that:
\beq
f_{\gamma\circ\beta\circ\alpha}={\cal R}_{\gamma\circ\beta\circ\alpha}\circ
T_{\beta\circ\alpha}
\eeq
If we perform this measurement on $S_\beta$ and we obtain an outcome state
$\phi_\beta$, then the proper state of $S_\gamma$ (for all $\gamma\not=\alpha,\beta$) changes to
$f_{\gamma\circ\beta\circ\alpha}(\phi_\beta)$. Clearly, we can proceed with this procedure for some
more consecutive measurements. 
After we have performed consecutive measurements on $S_\alpha$,
\ldots,$S_\kappa$, they are in the states respectively denoted as $\phi_\alpha$,
\ldots,$\phi_\kappa$, and the proper state of every not yet measured individual entity $S_\lambda$ is
given by
$\omega_{\lambda\circ\kappa\circ\ldots\circ\alpha}=
f_{\lambda\circ\kappa\circ\ldots\circ\alpha}(\phi_\kappa)$.
If we perform a next measurement on $S_\lambda$, the proper state of a not yet measured individual
entity
$S_\mu$ changes to
$f_{\mu\circ\lambda\circ\kappa\circ\ldots\circ\alpha}(\phi_\lambda)$, where $\phi_\lambda$ is the outcome
state of $S_\lambda$ and
\beq\label{eq.last.HC}
f_{\mu\circ\lambda\circ\kappa\circ\ldots\circ\alpha}:{\cal H}_\lambda\rightarrow\bar{\cal H}_\mu
\eeq   
is defined in an analogous way as
$f_{\gamma\circ\beta\circ\alpha}$, but now by considering a biorthogonal decomposition of 
$T_{\kappa\circ\ldots\circ\alpha}(\phi_\kappa)$ in stead of
$T_{\alpha}(\phi_\alpha)$.
We can summarize all this in the following definition:
\bdf\label{defrep}
If $S$ is a compound quantum system consisting of a finite collection of individual entities
$\{S_\nu\}_\nu$, and such that $S$ is described by $\Psi_S\in\otimes_\nu{\cal H}_\nu$, then
we can introduce a hidden correlation representation for $S$:
{\bf 1.} Initially, every $S_\alpha$ is in a proper state $\omega_\alpha={\cal R}_\alpha(\Psi_S)$.
{\bf 2.} If $S_\lambda$ takes the state
$\phi_\lambda$ due to a measurement on it after we have already performed
measurements on
$S_\alpha,\ldots,S_\kappa$, 
then the proper state of every not yet measured individual entity $S_\mu$ changes to 
$f_{\mu\circ\lambda\circ\kappa\circ\ldots\circ\alpha}(\phi_\lambda)=
{\cal R}_{\mu\circ\lambda\circ\kappa\circ\ldots\circ\alpha}\circ
T_{\lambda\circ\kappa\circ\ldots\circ\alpha}(\phi_\lambda)$.
\edf

\subsection{Correctness of the representation.}

Now we will prove that our representation respects the specific probabilistic nature of the quantum
description of a compound system.
\bth\label{th.correct}
The probabilities that we obtain in the
representation introduced in definition \ref{defrep} are the quantum
probabilities.   
\eth
\bpf
Suppose that the quantum system $S$ consists of the individual entities $\{S_\nu\}_\nu$, that $S$ is
described by
$\Psi_S\in\otimes_\nu{\cal H}_\nu$ and that we measure these individual
entities one by one according to the ordering $\alpha,\beta,\gamma,\ldots,\xi,\upsilon,\zeta$. 
First we calculate, according to the rules
of quantum mechanics, the probability to obtain an outcome state
$\otimes_\nu\phi_\nu$ in a measurement that has
$\otimes_\nu\phi_\nu$ as an eigenvector, taking into account the existence of
representations as a biorthogonal decomposition for vectors described in the tensor product of two
Hilbert spaces (to distinguish between the indices of the different bases corresponding with the
different Hilbert spaces in $\{ {\cal H}_\nu\}_\nu$ we introduce a subscript in the indices):
\beqa
\hspace{-0.6cm}P_{QM}\bigl(\Psi_S\bigm|\phi_\alpha\otimes\ldots\otimes\phi_\zeta\bigr)
&=&|\langle\Psi_S|\phi_\alpha\otimes\ldots\otimes\phi_\zeta\rangle|^2\\
&=&|\langle\sum_{i_\alpha}\langle\psi_{\alpha,{i_\alpha}}\otimes\tilde{\psi}_{\alpha,{i_\alpha}}|
\Psi_S\rangle\psi_{\alpha,{i_\alpha}}\otimes\tilde{\psi}_{\alpha,{i_\alpha}}|
\phi_\alpha\otimes\ldots\otimes\phi_\zeta\rangle|^2\\
&=&|\sum_{i_\alpha}\overline{\langle\psi_{\alpha,{i_\alpha}}\otimes\tilde{\psi}_{\alpha,{i_\alpha}}|
\Psi_S\rangle}\langle\psi_{\alpha,{i_\alpha}}\otimes\tilde{\psi}_{\alpha,{i_\alpha}}|
\phi_\alpha\otimes\ldots\otimes\phi_\zeta\rangle|^2\\
&=&|\sum_{i_\alpha}
\overline{\langle\psi_{\alpha,{i_\alpha}}\otimes\tilde{\psi}_{\alpha,{i_\alpha}}|\Psi_S\rangle}
\langle\psi_{\alpha,{i_\alpha}}|\phi_\alpha\rangle
\langle\tilde{\psi}_{\alpha,{i_\alpha}}|\phi_\beta\otimes\ldots\otimes\phi_\zeta\rangle|^2\\
&=&|\langle\sum_{i_\alpha}\langle\psi_{\alpha,{i_\alpha}}\otimes\tilde{\psi}_{\alpha,{i_\alpha}}|
\Psi_S\rangle\langle\phi_\alpha|\psi_{\alpha,{i_\alpha}}\rangle\tilde{\psi}_{\alpha,{i_\alpha}}|
\phi_\beta\otimes\ldots\otimes\phi_\zeta\rangle|^2
\eeqa
For $\sum_{i_\alpha}\langle\psi_{\alpha,{i_\alpha}}\otimes\tilde{\psi}_{\alpha,{i_\alpha}}|
\Psi_S\rangle\langle\phi_\alpha|\psi_{\alpha,{i_\alpha}}\rangle\tilde{\psi}_{\alpha,{i_\alpha}}$ we have
the following biorthogonal decomposition in
${\cal H}_\beta\otimes({\cal H}_\gamma\otimes\ldots\otimes{\cal H}_\zeta)$: 
\beqa
\sum_{i_\beta}\langle\psi_{\beta,{i_\beta}}\otimes\tilde{\psi}_{\beta,{i_\beta}}|
\sum_{i_\alpha}\langle\psi_{\alpha,{i_\alpha}}\otimes\tilde{\psi}_{\alpha,{i_\alpha}}|
\Psi_S\rangle\langle\phi_\alpha|\psi_{\alpha,{i_\alpha}}\rangle\tilde{\psi}_{\alpha,{i_\alpha}}\rangle
\psi_{\beta,{i_\beta}}\otimes\tilde{\psi}_{\beta,{i_\beta}}
\eeqa
thus one finds that 
$P_{QM}\bigl(\Psi_S\bigm|\phi_\alpha\otimes\ldots\otimes\phi_\zeta\bigr)$ is given by:
\beqa
&&\hspace{-1.4cm}
|\langle\sum_{i_\beta}\langle\psi_{\beta,{i_\beta}}\otimes\tilde{\psi}_{\beta,{i_\beta}}|
\sum_{i_\alpha}\langle\psi_{\alpha,{i_\alpha}}\otimes\tilde{\psi}_{\alpha,{i_\alpha}}|
\Psi_S\rangle\langle\phi_\alpha|\psi_{\alpha,{i_\alpha}}\rangle\tilde{\psi}_{\alpha,{i_\alpha}}\rangle
\psi_{\beta,{i_\beta}}\otimes\tilde{\psi}_{\beta,{i_\beta}}|
\phi_\beta\otimes\ldots\otimes\phi_\zeta\rangle|^2
\\
&&\hspace{-0.4cm}=
|\sum_{i_\alpha}\sum_{i_\beta}
\overline{\langle\psi_{\alpha,{i_\alpha}}\otimes\tilde{\psi}_{\alpha,{i_\alpha}}|\Psi_S\rangle
\langle\phi_\alpha|\psi_{\alpha,{i_\alpha}}\rangle}
\langle\psi_{\beta,{i_\beta}}\otimes\tilde{\psi}_{\beta,{i_\beta}}|
\tilde{\psi}_{\alpha,{i_\alpha}}\rangle
\langle\psi_{\beta,{i_\beta}}\otimes\tilde{\psi}_{\beta,{i_\beta}}|
\phi_\beta\otimes\ldots\otimes\phi_\zeta\rangle|^2
\\
&&\hspace{-0.4cm}=
|\sum_{i_\alpha}\sum_{i_\beta}
\overline{\langle\psi_{\alpha,{i_\alpha}}\otimes\tilde{\psi}_{\alpha,{i_\alpha}}|\Psi_S\rangle
\langle\phi_\alpha|\psi_{\alpha,{i_\alpha}}\rangle
\langle\psi_{\beta,{i_\beta}}\otimes\tilde{\psi}_{\beta,{i_\beta}}|
\tilde{\psi}_{\alpha,{i_\alpha}}\rangle}
\langle\psi_{\beta,{i_\beta}}|\phi_\beta\rangle\langle\tilde{\psi}_{\beta,{i_\beta}}|
\phi_\gamma\otimes\ldots\otimes\phi_\zeta\rangle|^2
\\
&&\hspace{-0.4cm}=
|\sum_{i_\alpha}
\overline{\langle\psi_{\alpha,{i_\alpha}}\otimes\tilde{\psi}_{\alpha,{i_\alpha}}|\Psi_S\rangle
\langle\phi_\alpha|\psi_{\alpha,{i_\alpha}}\rangle}
\langle\sum_{i_\beta}\langle\psi_{\beta,{i_\beta}}\otimes\tilde{\psi}_{\beta,{i_\beta}}|
\tilde{\psi}_{\alpha,{i_\alpha}}\rangle
\langle\phi_\beta|\psi_{\beta,{i_\beta}}\rangle\tilde{\psi}_{\beta,{i_\beta}}|
\phi_\gamma\otimes\ldots\otimes\phi_\zeta\rangle|^2
\eeqa 
Also for $\sum_{i_\beta}\langle\psi_{\beta,{i_\beta}}\otimes\tilde{\psi}_{\beta,{i_\beta}}|
\tilde{\psi}_{\alpha,{i_\alpha}}\rangle
\langle\phi_\beta|\psi_{\beta,{i_\beta}}\rangle\tilde{\psi}_{\beta,{i_\beta}}$ we can consider a
biorthogonal decomposition in the base
$\{\psi_{\gamma,{i_\gamma}}\otimes\tilde{\psi}_{\gamma,{i_\gamma}}\}_{i_\gamma}$.  If we proceed along
the same lines we obtain (we can omit the complex conjugation since it applies to all factors of
which we take the norm):
\beqa
&&\hspace{-1.1cm}
|\sum_{i_\alpha}\sum_{i_\beta}\ldots\sum_{i_\upsilon}\langle\psi_{\alpha,{i_\alpha}}\otimes\tilde{\psi}_{\alpha,{i_\alpha}}|\Psi_S\rangle
\langle\psi_{\beta,{i_\beta}}\otimes\tilde{\psi}_{\beta,{i_\beta}}|
\tilde{\psi}_{\alpha,{i_\alpha}}\rangle
\ldots
\langle\psi_{\upsilon,{i_\upsilon}}
\otimes\tilde{\psi}_{\upsilon,{i_\upsilon}}|
\tilde{\psi}_{\xi,{i_\xi}}\rangle
\\
&&\hspace{8.8cm}
\langle\psi_{\alpha,{i_\alpha}}|\phi_\alpha\rangle
\langle\psi_{\beta,{i_\beta}}|\phi_\beta\rangle
\ldots
\langle\tilde{\psi}_{\upsilon,{i_\upsilon}}|\phi_\zeta\rangle|^2
\eeqa
Now we calculate the probabilities in our representation. Let us use $P(\psi\rightarrow\phi)$ as a
notation for the probability of the transition of $\psi$ into $\phi$.
The probability to obtain an
outcome $\phi_\alpha\otimes\ldots\otimes\phi_\zeta$ for the measurement on $S$ is given by the
product: 
\beqa
P\bigl(\omega_\alpha\rightarrow\phi_\alpha\bigr)
P\bigl(f_{\beta\circ\alpha}(\phi_\alpha)\rightarrow\phi_\beta\bigr)
\ldots
P\bigl(f_{\zeta\circ\ldots\circ\alpha}(\phi_\upsilon)\rightarrow\phi_\zeta\bigr)
\eeqa
(although the initial proper state for every next
measurement depends on the outcome state of the previous one, the measurements themselves
are independent).  For the first factor we have:
\beqa
P\bigl(\omega_\alpha\rightarrow\phi_\alpha\bigr)
&=&
\sum_{i_\alpha}
|\langle\psi_{\alpha,{i_\alpha}}\otimes\tilde{\psi}_{\alpha,{i_\alpha}}|\Psi_S\rangle|^2 
P\bigl(\psi_{\alpha,{i_\alpha}}\rightarrow\phi_\alpha\bigr)\\
&=&
\sum_{i_\alpha}
|\langle\psi_{\alpha,{i_\alpha}}\otimes\tilde{\psi}_{\alpha,{i_\alpha}}|\Psi_S\rangle|^2 
|\langle\psi_{\alpha,{i_\alpha}}|\phi_\alpha\rangle|^2\\
&=& 
N(\phi_\alpha)^2
\eeqa
Thus, $N(\phi_\alpha)^2$ is the probability to obtain an outcome state $\phi_\alpha$ when we are in
proper state $\omega_\alpha$, which is itself is determined by $\Psi_S$. Since:
\beqa
f_{\beta\circ\alpha}(\phi_\alpha)&=&{\cal R}_{\beta\circ\alpha}\bigl(T_\alpha(\phi_\alpha)\bigr)
\eeqa
we have (according to the definition of ${\cal R}_{\beta\circ\alpha}$): 
\beqa
P\bigl(f_{\beta\circ\alpha}(\phi_\alpha)\rightarrow\phi_\beta\bigr)
&=&
\sum_{i_\beta}
|\langle\psi_{\beta,i_\beta}\otimes\tilde{\psi}_{\beta,i_\beta}|
T_\alpha(\phi_\alpha)\rangle|^2
P\bigl(\psi_{\beta,{i_\beta}}\rightarrow\phi_\beta\bigr)
\\
&=&
\sum_{i_\beta}
|\langle\psi_{\beta,i_\beta}\otimes\tilde{\psi}_{\beta,i_\beta}|
T_\alpha(\phi_\alpha)\rangle|^2
|\langle\psi_{\beta,i_\beta}|\phi_\beta\rangle|^2
\\
&=&
{N(\phi_\beta)^2}
\eeqa
where we applied the same biorthogonal decomposition as in
the calculation of the quantum case, since $T_\alpha(\phi_\alpha)$ equals
$\sum_{i_\alpha}\langle\psi_{\alpha,i_\alpha}\otimes\tilde{\psi}_{\alpha,i_\alpha}|\Psi_S\rangle
\langle\phi|\psi_{\alpha,i_\alpha}\rangle\tilde{\psi}_{\alpha,i_\alpha}$ up to a the constant
$N(\phi_\alpha)$, and  where $N(\phi_\beta)$ is defined by eq.(\ref{eq:Tba}).  This gives:
\beqa
&P\bigl((\omega_\alpha,\ldots,\omega_\zeta)\rightarrow(\phi_\alpha\,\ldots,\phi_\zeta)\bigr)=
N(\phi_\zeta)^2\ldots N(\phi_\upsilon)^2
P\bigl(f_{\beta\circ\ldots\circ\zeta}(\phi_\upsilon)\rightarrow\phi_\zeta\bigr)&
\eeqa
For the last factor we have that $T_\upsilon(\phi_\upsilon)\in {\cal H}_\zeta$ and thus:
\beqa
P\bigl(f_{\zeta\circ\ldots\circ\alpha}(\phi_\upsilon)\rightarrow\phi_\zeta\bigr)
&=&
P\bigl(T_\upsilon(\phi_\upsilon)\rightarrow\phi_\zeta\bigr)
\\
&=&
|\langle\phi_\zeta|T_\upsilon(\phi_\upsilon)|^2
\\
&=&
|\langle\phi_\zeta|
{1\over N(\phi_\upsilon)}
\sum_{i_\upsilon}\langle\psi_{\upsilon,i_\upsilon}\otimes\tilde{\psi}_{\upsilon,i_\upsilon}|
T_\xi(\phi_\xi)\rangle
\langle\phi_\upsilon|\psi_{\upsilon,i_\upsilon}\rangle\tilde{\psi}_{\upsilon,i_\upsilon}
\rangle|^2
\\
&=&
{1\over N(\phi_\upsilon)^2}
|\sum_{i_\upsilon}\langle\psi_{\upsilon,i_\upsilon}\otimes\tilde{\psi}_{\upsilon,i_\upsilon}|
T_\xi(\phi_\xi)\rangle
\langle\phi_\upsilon|\psi_{\upsilon,i_\upsilon}\rangle\langle\phi_\zeta|
\tilde{\psi}_{\upsilon,i_\upsilon}
\rangle|^2
\eeqa
A consecutive substitution of $T_\xi(\phi_\xi),\ldots,T_\alpha(\phi_\alpha)$ in 
$P\bigl(f_{\zeta\circ\ldots\circ\alpha}(\phi_\upsilon)\rightarrow\phi_\zeta\bigr)$ gives:
\beqa
&&\hspace{-0.5cm}
{1\over N(\phi_\alpha)^2}\ldots{1\over N(\phi_\upsilon)^2}
|\sum_{i_\alpha}\sum_{i_\beta}\ldots\sum_{i_\upsilon}
\langle\psi_{\alpha,{i_\alpha}}\otimes\tilde{\psi}_{\alpha,{i_\alpha}}|\Psi_S\rangle
\langle\psi_{\beta,{i_\beta}}\otimes\tilde{\psi}_{\beta,{i_\beta}}|
\tilde{\psi}_{\alpha,{i_\alpha}}\rangle
\ldots
\langle\psi_{\upsilon,{i_\upsilon}}
\otimes\tilde{\psi}_{\upsilon,{i_\upsilon}}|
\tilde{\psi}_{\xi,{i_\xi}}\rangle
\\
&&\hspace{9.8cm}
\langle\psi_{\alpha,{i_\alpha}}|\phi_\alpha\rangle
\langle\psi_{\beta,{i_\beta}}|\phi_\beta\rangle
\ldots
\langle\tilde{\psi}_{\upsilon,{i_\upsilon}}|\phi_\zeta\rangle|^2
\eeqa
and thus, we find for
$P\bigl((\omega_\alpha,\ldots,\omega_\zeta)\rightarrow(\phi_\alpha\,\ldots,\phi_\zeta)\bigr)$ the same
expression as we have found in the quantum calculation.
\epf

\subsection{Uniqueness of the representation.}\label{secunique}

Now we will prove that the representation of
definition 2 is the only possible one that fulfills definition 1.  
\bth
The representation introduced in section 2 is unique, i.e., there exist no other hidden correlation
representations for compound quantum systems described in the tensor product of a finite number of
Hilbert spaces.
\eth
\bpf
We have to prove that the density matrix representations $\{\omega_\eta\}_\eta$ for the proper
states, as well as the transitions of them induced by measurements on individual
entities are uniquely determined by $\Psi_S$. Suppose that there exist other 
$\{\omega_\eta'\}_\eta$ which fulfill definition 1 (initially we allow the state transitions to
depend on other variables than the final state of the performed measurement). 
For all $\nu\not=\alpha$, let $\{\phi_{\nu,i_\nu}\}_{i_\nu}$ be an orthonormal base. We have for
every
$\phi_\alpha\in {\cal H}_\alpha$:
\beqa
&&\hspace{-1.5cm}
\sum_{\i_\beta}\ldots\sum_{i_\zeta}
P_{QM}\bigl(\Psi_S\bigm|\phi_\alpha\otimes\phi_{\beta,i_\beta}\otimes\ldots\otimes\phi_{\zeta,i_\zeta}\bigr)
\\
&&\hspace{-0.5cm}=
\sum_{\i_\beta}\ldots\sum_{i_\zeta}
P\bigl(\omega_\alpha'\rightarrow\phi_\alpha\bigr)
P\bigl(f_{\beta\circ\alpha}'(\phi_\alpha,\ldots)\rightarrow\phi_{\beta,i_\beta}\bigr)
\ldots
P\bigl(f_{\zeta\circ\ldots\circ\alpha}'(\phi_{\upsilon,i_\upsilon},\ldots)\rightarrow\phi_{\zeta,i_\zeta}\bigr)
\\
&&\hspace{-0.5cm}=
P\bigl(\omega_\alpha'\rightarrow\phi_\alpha\bigr)
\Bigl(\sum_{\i_\beta}P\bigl(f_{\beta\circ\alpha}'(\phi_\alpha,\ldots)\rightarrow
\phi_{\beta,i_\beta}\bigr)\Bigr)
\ldots
\Bigl(\sum_{i_\zeta}P\bigl(f_{\zeta\circ\ldots\circ\alpha}'(\phi_{\upsilon,i_\upsilon},\ldots)\rightarrow
\phi_{\zeta,i_\zeta}\bigr)\Bigr)
\\
&&\hspace{-0.5cm}=
P_{QM}\bigl(\omega_\alpha'|\phi_\alpha\bigr)
\eeqa
since all sums between brackets are equal to one (as a consequence of the normalization of the
Hilbert in-product). Due to the independence of $\sum_{\i_\beta}\ldots\sum_{i_\zeta}
P_{QM}\bigl(\Psi_S\bigm|\phi_\alpha\otimes\phi_{\beta,i_\beta}\otimes\ldots\otimes\phi_{\zeta,i_\zeta}\bigr)
$ on $\omega_\alpha'$, we have for every $\phi_\alpha\in {\cal
H}_\alpha$ that
$P_{QM}\bigl(\omega_\alpha|\phi_\alpha\bigr)=P_{QM}\bigl(\omega_\alpha'|\phi_\alpha\bigr)$, and thus,
$\omega_\alpha=\omega_\alpha'$, since this expression implicitly defines a density matrix
as a positive normalized measure, $\sigma$-additive on mutual orthogonal subspaces, on the subspaces
of the Hilbert space. In an analogous way we prove the same for the other
proper states in
$\{\omega_\nu\}_\nu$, by considering the first measurement performed on them.   In a rather similar way we also prove the
uniqueness of the imposed transitions of these proper states. We illustrate this for the proper state
transition of
$S_\beta$  induced by a measurement on $S_\alpha$, i.e., $\omega_\beta$
changes into
$f_{\beta\circ\alpha}'(\phi_\alpha,\ldots)$ due to a transition of $\omega_\alpha$ into
$\phi_\alpha$:
\beqa
&&\hspace{-1.2cm}
\sum_{\i_\gamma}\ldots\sum_{i_\zeta}
P_{QM}\bigl(\Psi_S\bigm|\phi_\alpha\otimes\phi_{\beta}\otimes\phi_{\gamma,i_\gamma}\otimes
\ldots\otimes\phi_{\zeta,i_\zeta}\bigr)
\\
&&\hspace{-0.1cm}=
P\bigl(\omega_\alpha\rightarrow\phi_\alpha\bigr)
P\bigl(f_{\beta\circ\alpha}'(\phi_\alpha,\ldots)\rightarrow\phi_\beta\bigr)
\\
&&\hspace{3.2cm}
\Bigl(\sum_{\i_\gamma}P\bigl(f_{\gamma\circ\beta\circ\alpha}'(\phi_\gamma,\ldots)\rightarrow
\phi_{\gamma,i_\gamma}\bigr)\Bigr)
\ldots
\Bigl(\sum_{i_\zeta}P\bigl(f_{\zeta\circ\ldots\circ\alpha}'(\phi_{\upsilon,i_\upsilon},\ldots)\rightarrow
\phi_{\zeta,i_\zeta}\bigr)\Bigr)
\\
&&\hspace{-0.1cm}=
P\bigl(\omega_\alpha\rightarrow\phi_\alpha\bigr)
P_{QM}\bigl(f_{\beta\circ\alpha}'(\phi_\alpha,\ldots)|\phi_\beta\bigr)
\eeqa
The state transition of $S_\beta$ from $\omega_\beta$ into 
$f_{\beta\circ\alpha}'(\phi_\alpha,\ldots)$ happens after a transition of the proper state
$\omega_\alpha$ into
$\phi_\alpha$ and thus $P\bigl(\omega_\alpha\rightarrow\phi_\alpha\bigr)\not=0$. Thus we have:
\beq
P_{QM}\bigl(f_{\beta\circ\alpha}'(\phi_\alpha,\ldots)|\phi_\beta\bigr)
=
{\sum_{\i_\gamma}\ldots\sum_{i_\zeta}
P_{QM}\bigl(\Psi_S\bigm|\phi_\alpha\otimes\phi_{\beta}\otimes\phi_{\gamma,i_\gamma}\otimes
\ldots\otimes\phi_{\zeta,i_\zeta}\bigr)
\over
P\bigl(\omega_\alpha\rightarrow\phi_\alpha\bigr)}
\eeq
for every $\phi_\beta\in {\cal H}_\beta$. As a consequence,
$f_{\beta\circ\alpha}'(\phi_\alpha,\ldots)$ is equal to $f_{\beta\circ\alpha}(\phi_\alpha)$.
\epf

\section{Summary and conclusion.}

To conclude, for all states described in the tensor product of Hilbert spaces there exists a unique
representation of the kind we have defined in this paper, i.e., a representation as a
collection of individual entities each in a proper state, and such that a measurement on one of the
entities induces a transition of the proper states of the other ones. 
Due to the uniqueness theorem, the representation introduced in this paper gives a
definite characterization of the {\it new kind of compoundness} that can be described in a tensor
product and not in a cartesian product.  The approach can be embedded in a very natural way within
Aerts' creation-discovery approach and Piron's property-approach if we introduce the notions of
{\it proper state}, {\it individual entity} and {\it hard and soft acts of creation}.  In
particular does this approach deliver the appropriate tools to express {\it compoundness} of
physical systems within the hidden measurement formalism.  From the explicit construction of such
a hidden measurement representation for a compound system it follows that our approach can be
considered as a theory of individuals relative to the quantum description in a tensor product of
Hilbert spaces.  This is due to the fact that in our approach, the interaction of an individual
entity with the other individual entities is threaded in the same way as the interaction with the
measurement context.  We also delivered an argument why our {\it theory of individuals}
might be considered as preferable: the choice of the order in which we perform the measurements on the
individual entities in the compound system should be an ingredient of the description of the context
and not of the entity itself. 
 
In fact, in this paper we only studied the situation in which after the measurements on the individual
entities, these entities are separated due to a hard act of creation.  In [16] we introduce a
representation for a spin-$S$ entity as a compound system consisting of $2S$
individual spin-${1\over 2}$ entities.  Although in that paper we focus more on the formal
ingredients of the representation, and in particular on the connection with Majorana's
representation, the representation itself delivers an example within quantum theory where a hidden
correlation representation is applied in order to describe a situation where there are no hard acts
of creation: the individual entities in the compound systems are not separated
by the measurements since the spin-$S$ entity is still a spin-$S$ entity after the measurement,
i.e., there is no change of the state space.    
 
Of course, a lot of questions, as well on the formal as on the philosophical level remain
unanswered.  However, we think that the explicit nature of the representations for compound systems
introduced in this paper allows it to be explicitly confronted with some other points of view, and
to contribute in a constructive way to some of the remaining problems in the understanding of
physics.  For example: a somewhat metaphorical characterization of symmetric and
anti-symmetric superpositions has
been given in \cite{coe96}, and from this might result an answer to the question why we only encounter
them in nature; a more general and explicit answer might result from the
representation introduced in this paper, by an explicitation in it of the ideas launched in
\cite{coe96}.

\section{Appendix: independence of the choice of the biorthogonal decomposition.}

In this appendix we show the well-definedness of our representation, i.e., it does not depend on the
choice of the biorthogonal decomposition if
there exist more than one.  
Let $\sum_i a_i\psi_i\otimes\tilde{\psi}_i$ be a biorthogonal decomposition of $\Psi$ and suppose
that
$\sum_i a_i'\psi_i'\otimes\tilde{\psi}_i'$ is a second one.  Since the vectors in
$\{\psi_i\}_i$, $\{\tilde{\psi}_i\}_i$, $\{\psi_i'\}_i$ and $\{\tilde{\psi}_i'\}_i$ that appear explicitly
in the biorthogonal decompositions can always be extended to orthonormal bases of equal dimensional
Hilbert spaces, there exist two unitary matrices $\{U_{i,j}\}_{i,j}$ and $\{V_{i,j}\}_{i,j}$ such that
for all $i$ we have $\psi_i'=\sum_j U_{i,j}\psi_j$ and $\tilde{\psi}_i'=\sum_j V_{i,j}\tilde{\psi}_j$. Since both
biorthogonal decomposition are representations of the same state we have for all $i$ and $j$:
\beq
\langle\psi_i\otimes\tilde{\psi}_j|\sum_k a_k'\psi_k'\otimes\tilde{\psi}_k'\rangle
=\langle\psi_i\otimes\psi_j|\sum_k a_k\psi_k\otimes\tilde{\psi}_k\rangle\\
=\delta_{i,j}a_i
\eeq
Since,
\beq
\langle\psi_i\otimes\tilde{\psi}_i|\sum_k a_k'\psi_k'\otimes\tilde{\psi}_k'\rangle
=\sum_k a_k'U_{k,i}V_{k,j}
\eeq
we find:
\beq\label{uniqueUV}
\sum_k a_k'U_{k,i}V_{k,j}=\delta_{i,j}a_i
\eeq
As a consequence, we also have:
\beq
\sum_j\sum_k a_k'U_{k,i}V_{k,j}\overline{\sum_l a_l'U_{l,m}V_{l,j}}
=|a_i|^2\delta_{i,m}
\eeq
Since,
\beq
\sum_j\sum_k a_k'U_{k,i}V_{k,j}\overline{\sum_l a_l'U_{l,m}V_{l,j}}
=\sum_k|a_k'|^2U_{k,i}\bar U_{k,m}
\eeq
we obtain:
\beq\label{uniqueUU}
\sum_k|a_k'|^2U_{k,i}\bar U_{k,j}=|a_i|^2\delta_{i,j}
\eeq
For the second biorthogonal decomposition we find:
\beqa
{1\over N'(\phi)}\sum_i a_i\langle\phi|\psi_i'\rangle\tilde{\psi}_i'
={1\over N'(\phi)}\sum_{i,j,k}a_i U_{i,j}
V_{i,k}\langle\phi|\psi_i\rangle\tilde{\psi}_k
={1\over N'(\phi)}\sum_j a_j\langle\phi|\psi_i\rangle\tilde{\psi}_j
\eeqa
by applying eq.(\ref{uniqueUV}), and where:
\beqa
N'(\phi)=\sqrt{\sum_i|a_i'|^2 |\langle\sum_j U_{i,j}\psi_j|\phi\rangle|^2}
=\sqrt{\sum_j|a_j|^2|\langle\psi_j|\phi\rangle|^2=N(\phi)}
\eeqa
by applying eq.(\ref{uniqueUU})  Along the same lines we can also prove that the choice of the
biorthogonal decomposition doesn't influence $T_{\zeta\circ\ldots\circ\alpha}$.  Since $N(\phi)$ is
equal to the probability of a transition to a state $\phi$ when we are in a proper state
$\omega_{\xi\circ\zeta\circ\ldots\circ\alpha}$, we have also proved that both biorthogonal
decompositions always determine the same map
${\cal R}_{\xi\circ\zeta\circ\ldots\circ\alpha}$.  
 
\section{Acknowledgments.}    

We also thank
D. Dieks, F. Valckenborgh and P. Vermaas for reading this paper and for discussing some
interesting questions pointed out by the referee. 
We thank the referee for insisting on a more elaborated philosophical framework for the representation
introduced in this paper.


\begin{thebibliography}{99}

\bibitem{aer81}  
\noindent {\bf D. Aerts}, {\it The One and the Many}, Doctoral Dissertation,
Free University of Brussels (1981); {\it Found. Phys.} {\bf 12}, 1131 (1982).
\vspace{-0.2cm}

\bibitem{aer86}  
\smallskip
\noindent {\bf D. Aerts}, {\it J. Math. Phys.} {\bf 27}, 202 (1986).
\vspace{-0.2cm}  

\bibitem{aer91}  
\smallskip
\noindent {\bf D. Aerts}, {\it Helv. Phys. Acta} {\bf 64}, 1 (1991).
\vspace{-0.2cm}

\bibitem{aer94}  
\smallskip
\noindent {\bf D. Aerts}, {\it Found. Phys.} {\bf 24}, 1227 (1994). 
\vspace{-0.2cm}

\bibitem{aer97}   
\smallskip
\noindent {\bf D. Aerts}, 'The Entity and Modern Physics: The Creation-Discovery View of Reality',
in {\em Interpreting Bodies: Classical and Quantum Objects in Modern Physics}, pp.~223-257, ed. E.
Castellani, Princeton University Press, New Jersey (1998).
\vspace{-0.2cm}

\bibitem{aercoe}   
\smallskip
\noindent {\bf D. Aerts and B. Coecke}, 'The Creation-Discovery-View : Towards a Possible
Explanation of Quantum Reality', in {\it Language, Quantum, Music}, pp.~105--116, eds. M.L.
Dalla-Ciara {\sl et al.},
Kluwer Academic Publishers, Dordrecht (1999).
\vspace{-0.2cm}

\bibitem{bell64}  
\smallskip
\noindent {\bf J. Bell}, {\it Physics} {\bf 1}, 195 (1964).
\vspace{-0.2cm}

\bibitem{bel}  
\smallskip
\noindent {\bf E.G. Beltrametti and G. Cassinelli}, {\it The Logic of Quantum
Mechanics}, Addison-Wesley Publishing Company, London (1981).
\vspace{-0.2cm}

\bibitem{bit82}  
\smallskip
\noindent {\bf E. Bitsakis}, {\it Scientia} {\bf 117}, 561 (1982).
\vspace{-0.2cm}

\bibitem{cat91}  
\smallskip
\noindent {\bf G. Cattaneo and G. Nistico}, {\it Int. J. Theor. Phys.} {\bf 30}, 1293 (1991). 
\vspace{-0.2cm}

\bibitem{cat93}  
\smallskip
\noindent {\bf G. Cattaneo and G. Nistico}, {\it Int. J. Theor. Phys.} {\bf 32}, 407 (1993).
\vspace{-0.2cm}

\bibitem{cla}   
\smallskip
\noindent {\bf J.F. Clauser and A. Shimony}, {\it Rep. Prog. Phys.} {\bf 41}, 1881 (1978).
\vspace{-0.2cm}

\bibitem{cli95}   
\smallskip
\noindent {\bf R. Clifton}, {\it British J. Phil. Sc.} {\bf 46}, 33 (1995).
\vspace{-0.2cm}

\bibitem{coe95}  
\smallskip
\noindent {\bf B. Coecke}, {\it Found. Phys. Lett.} {\bf 8}, 437 (1995).
\vspace{-0.2cm}

\bibitem{coe96}  
\smallskip
\noindent {\bf B. Coecke}, {\it Int. J. Theor. Phys.} {\bf 35}, 1217 (1996).
\vspace{-0.2cm}

\bibitem{coe96b}  
\smallskip
\noindent {\bf B. Coecke}, {\it Hidden Measurement Systems}, Doctoral Dissertation,
Free University of Brussels (1996); {\it Helv. Phys. Acta} {\bf 70}, 442 (1997);
{\it Helv. Phys. Acta} {\bf 70}, 462 (1997)\,; arXiv: quant-ph/0008061 \& 0008062\,.
\vspace{-0.2cm}

\bibitem{spinN}   
\smallskip
\noindent {\bf B. Coecke}, {\it Found. Phys.} {\bf 28}, 1347 (1998).
\vspace{-0.2cm}

\bibitem{coe2000}    
\smallskip
\noindent {\bf B. Coecke}, {\it Int. J. Theor. Phys.} {\bf 39}, 581 (2000)\,; arXiv: quant-ph/0008054.
\vspace{-0.2cm}

\bibitem{bofr}  
\smallskip
\noindent {\bf B. Coecke and F. Valckenborgh}, {\it Int. J. Theor. Phys.} {\bf 37}, 311 (1998).
\vspace{-0.2cm} 

\bibitem{esp}  
\smallskip
\noindent {\bf B. d'Espagnat}, {\it Conceptual Foundations of Quantum Mechanics}, W.A. Benjamin,
London (1976).
\vspace{-0.2cm}

\bibitem{fou84} 
\smallskip
\noindent {\bf D.J. Foulis and C.H. Randall}, {\it Found Phys.} {\bf 14}, 65 (1984).
\vspace{-0.2cm}

\bibitem{free72} 
\smallskip
\noindent {\bf S.J. Freedman and J.F. Clauser}, {\it Phys. Rev. Lett.} {\bf 28}, 938 (1972).
\vspace{-0.2cm}

\bibitem{gis81} 
\smallskip
\noindent {\bf N. Gisin and C. Piron}, {\it Lett. Math. Phys.} {\bf 5}, 379 (1981).
\vspace{-0.2cm}

\bibitem{glea}  
\smallskip
\noindent {\bf A.M. Gleason}, {\it J. Math. Mech.} {\bf 6}, 885 (1957).
\vspace{-0.2cm}

\bibitem{hea89}
\smallskip
\noindent {\bf R. Healey}, {\it The Philosophy of Quantum Mechanics}, 
Cambridge Univ. Press, Cambridge (1989).
\vspace{-0.2cm}

\bibitem{her1}   
\smallskip
\noindent {\bf F. Herbut and M. Vuji\~ci\'c}, {\it Ann. Phys.} {\bf 96}, 382 (1976).
\vspace{-0.2cm}

\bibitem{moo}  
\smallskip
\noindent {\bf D.J. Moore}, {\it Helv. Phys. Acta} {\bf 68}, 658 (1995). 
\vspace{-0.2cm}

\bibitem{moo97}  
\smallskip
\noindent {\bf D.J. Moore}, 
{\it Stud. Hist. Phil. Mod. Phys.} {\bf 30}, 61 (1999).
\vspace{-0.2cm} 

\bibitem{pau}  
\smallskip
\noindent {\bf W. Pauli}, {\it Die Allgemeinen Prinzipien der Wellenmechanic}, Handbuch der Physik
Vol. V, Part I, Springer-Verlag, Berlin (1958).
\vspace{-0.2cm}
 
\bibitem{pir}  
\smallskip
\noindent {\bf C. Piron}, {\it Foundations of Quantum Physics}, W.A. Benjamin,
London (1976).
\vspace{-0.2cm}

\bibitem{schm}  
\smallskip
\noindent {\bf E. Schmidt}, {\it Math. Ann.} {\bf 63}, 433 (1907).
\vspace{-0.2cm}

\bibitem{schro35}  
\smallskip
\noindent {\bf E. Schr\"odinger}, {\it Proc. Cambridge Philos. Soc.} {\bf 31}, 555 (1935).
\vspace{-0.2cm}

\bibitem{schro36}  
\smallskip
\noindent {\bf E. Schr\"odinger}, {\it Proc. Cambridge Philos. Soc.} {\bf 32}, 446 (1936).
\vspace{-0.2cm}

\bibitem{val2000}
\smallskip
\noindent {\bf F. Valckenborgh}, 'Operational Axiomatics and Compound Systems', in {\it Current Research in
Operational Quantum Logic: Algebras Categories, Languages}, eds. B. Coecke, D.J.
Moore and A. Wilce, Kluwer Academic Publishers, Dordrecht (2000).
\vspace{-0.2cm}

\bibitem{fra}   
\smallskip
\noindent {\bf B.C. van Fraassen}, {\it Quantum Mechanics}, Clarendon Press,
Oxford (1991).
\vspace{-0.2cm}

\bibitem{neu}   
\smallskip
\noindent {\bf J. von Neumann}, {\it The Mathematical Foundations of Quantum Mechanics},
Princeton University Press, Princeton (1955).
\vspace{-0.2cm}

\bibitem{her2}  
\smallskip
\noindent {\bf M. Vuji\~ci\'c and F. Herbut}, {\it J. Math. Phys.} {\bf 25}, 2253 (1984).
 
\end{thebibliography}
\end{document}